\documentclass{emulateapj}
\usepackage{ulem}
\usepackage{color}
\usepackage{url}

\shorttitle{HAeBes in the Quartet}

\shortauthors{Yasui et al.}

\begin{document}

\title{Herbig Ae/Be candidate stars in the innermost Galactic disk: 
Quartet cluster} 
%
%
%(Protoplanetary disks of ...(?))} 

%\altaffilmark{1}}
%
%\altaffiltext{1}{Based on data collected at Subaru Telescope, which
%is operated by the National Astronomical Observatory of Japan.}

%%% \author{Genevieve J. Graves\altaffilmark{1,2,3} \&
%%%   S. M. Faber\altaffilmark{1}}
%%% 
%%% \altaffiltext{1}{UCO/Lick Observatory, Department of Astronomy and
%%% Astrophysics, University of California, Santa Cruz, CA 95064, USA}
%%% \altaffiltext{2}{Department of Astronomy, University of California,
%%% Berkeley, CA 94720, USA; graves@astro.berkeley.edu}
%%% \altaffiltext{3}{Miller Fellow}

%Yasui, Kobayashi,Hamano, Kondo, Izumi, Tokunaga, Saito
\author{Chikako Yasui\altaffilmark{1,2}, Naoto
Kobayashi\altaffilmark{3,4,2}, Satoshi Hamano\altaffilmark{2}, Sohei
Kondo\altaffilmark{2}, Natsuko Izumi\altaffilmark{3,2}, Masao
Saito\altaffilmark{5,6}, and Alan T. Tokunaga\altaffilmark{7}} 
%, and Sohei Kondo\altaffilmark{2}}
% \footnote{Also at: Subaru
%Telescope, National Astronomical Observatory of Japan, 650 North A`ohoku
%Place, Hilo, HI 96720, USA.}}
%
%\altaffiltext{3}{Miller Fellow}
\altaffiltext{1}{Department of Astronomy, Graduate School of Science,
University of Tokyo, Bunkyo-ku, Tokyo 113-0033, Japan;
{ck.yasui@astron.s.u-tokyo.ac.jp}} 
%\email{ck.yasui@astron.s.u-tokyo.ac.jp}
%
%ALMA Project, National Astronomical Observatory of
%Japan, 2-21-1 Osawa, Mitaka, Tokyo 181-8588, Japan}

\altaffiltext{2}{Laboratory of Infrared High-resolution spectroscopy
(LIH), Koyama Astronomical Observatory, Kyoto Sangyo University,
Motoyama, Kamigamo, Kita-ku, Kyoto 603-8555, Japan}

\altaffiltext{3}{Institute of Astronomy, School of Science, University
of Tokyo, 2-21-1 Osawa, Mitaka, Tokyo 181-0015, Japan}
%\affil{Institute of Astronomy, School of Science, University of Tokyo,
%2-21-1 Osawa, Mitaka, Tokyo 181-0015, Japan}
%
%\email{ck$_-$yasui@ioa.s.u-tokyo.ac.jp}

\altaffiltext{4}{Kiso Observatory, Institute of Astronomy, School of
Science, University of Tokyo, 10762-30 Mitake, Kiso-machi, Kiso-gun,
Nagano 397-0101, Japan}

%\and

%\altaffiltext{4}{National Astronomical Observatory of Japan 2-21-1 
%Osawa, Mitaka, Tokyo, 181-8588, Japan}

\altaffiltext{5}{Nobeyama Radio Observatory, 462-2 Nobeyama,
Minamimaki-mura, Minamisaku-gun, Nagano 384-1305, Japan}

\altaffiltext{6}{The Graduate University of Advanced Studies,
(SOKENDAI), 2-21-1 Osawa, Mitaka, Tokyo 181-8588, Japan} 

%\author{Alan Tokunaga\altaffilmark{4,5}}
%\author{Alan T. Tokunaga}
\altaffiltext{7}{Institute for Astronomy, University of Hawaii, 2680 Woodlawn
Drive, Honolulu, HI 96822, USA}
%\affil{Institute for Astronomy, University of Hawaii, 2680 Woodlawn
%Drive, Honolulu, HI 96822, USA}

%\affil{Astronomical Institute, Tohoku University,
%Aramaki, Aoba, Sendai 980-8578, Japan}

%%%%%%%%%%%%%%%%%%%%%%%%%%%%%%%%%%%%%%%%%%%%%%%%%%%%%%%%%%%%%%%%%%%%%%%%%%%%%%%

\begin{abstract}

In order to investigate the Galactic-scale environmental effects 
on the evolution of protoplanetary
disks, we explored the near-infrared (NIR) disk fraction of the Quartet
cluster, which is a young cluster in the innermost Galactic disk at the
Galactocentric radius $R_g \sim 4$\,kpc.  Because this cluster has a
typical cluster mass of $\sim$$10^3$\,$M_\odot$ as opposed to very
massive clusters, which have been observed in previous studies
($>$$10^4$\,$M_\odot$), we can avoid intra-cluster effects such as
strong UV field from OB stars. 
Although the age of the Quartet is previously estimated to be 3--8\,Myr
old, we find that it is most likely $\sim$3--4.5\,Myr old. 
In moderately deep JHK images from the UKIDSS survey, we found eight
HAeBe candidates in the cluster, and performed K-band medium-resolution
($R \equiv \Delta \lambda / \lambda \sim 800$) spectroscopy for three of
them with the Subaru 8.2\,m telescope.  These are found to have both
Br$\gamma$ absorption lines as well as CO bandhead {\it emission},
suggesting that they are HAeBe stars with protoplanetary disks.
We estimated the intermediate-mass disk fraction (IMDF) to be
$\sim$25\,\% for the cluster, suggesting slightly higher IMDF compared
to those for young clusters in the solar neighborhood with similar
cluster age, although such conclusion should await future spectroscopic
study of all candidates of cluster members. 

\end{abstract}

% http://www.journals.uchicago.edu/page/apj/instruct.key.html


% (Galaxy:) open clusters and associations: individual (..., ...) 

%%%%%%%%%%%%%%%%%%%%%%%%%%%%%%%%%%%%%%%%%%%%%%%%%%%%%%%%%%%%%%%%%%%%%%%%%%%%%%%
%%%%%%%%%%%%%%%%%%  INTRODUCTION %%%%%%%%%%%%%%%%%%%%%%%%%%%%%%%%%%%%%%%%%%%%%
%%%%%%%%%%%%%%%%%%%%%%%%%%%%%%%%%%%%%%%%%%%%%%%%%%%%%%%%%%%%%%%%%%%%%%%%%%%%%%%

\section{INTRODUCTION} \label{sec:intro}

%\noindent
%\hspace{-5mm} {\bf Laboratory to see environmental dependence of star
%and planet formation}

Whether stars form in the similar way everywhere in different
physical/chemical environments or not is of great interest. Many groups
are studying the initial mass function (IMF) in various environments,
including the LMC and SMC, to find little difference (e.g.,
\citealt{Bastian2010}), which appears to suggest the universality of
star formation.  In the Galaxy, environments largely differ depending on
the Galactic radius, e.g., metallicity, gas density, perturbation by
spiral arms, which may offer an opportunity to study the environmental
dependence of star formation \citep{Kobayashi2008}.

Previously, we studied young clusters in the extreme outer Galaxy with
$R_g \gtrsim 18$\,kpc and suggested that the IMFs and star formation
efficiency (SFE) are not largely different from the solar neighborhood
\citep{{Yasui2006},{Yasui2008ASPC},{Yasui2008}} 
%(Yasui et al. 2006, 2008a, 2008b) 
except for the lifetime of protoplanetary disks, which was found to be
significantly shorter than in the solar neighborhood
\citep{{Yasui2009},{Yasui2010}}.  This might suggest that the
environment affects disk evolution and thus planet formation more than
star formation.

Star formation could also be largely affected by intra-cluster effects
e.g., stellar winds, radiation, and dynamical effects \citep{Adams2006}.
%\citep{{Adams2006}, {Stahler2013}}. 
Because such effects largely depend on the existence of massive stars,
the resultant star formation may also differ with the cluster mass
($M_{\rm cl}$) \citep{Harayama2008}.  For very massive clusters
($>$$10^4$\,$M_\odot$), the IMF may be different than the standard IMF
\citep{Bastian2010}.  The dependence of the disk lifetime on the cluster
mass and density is also suggested {\citep[e.g.,][]{{Stolte2010},
{Fang2013}}.
%e.g., \citealt{Stolte2010} and \citealt{Fang2013}}. 

%\vspace{1em}
%\noindent 
%\hspace{-5mm} 
%{\bf Target: the Quartet cluster} \\

As the next step for our study of the disk fraction in stellar clusters
as a function of Galactocentric radius \citep{{Yasui2009},{Yasui2010}},
we study the Quartet cluster that is a young cluster discovered by
% Messineo et al. (2009)
\citet{Messineo2009} from Spitzer/GLIMPSE images in the Galactic plane
at $(l,b) = (24.90^\circ, +0.12^\circ)$. 
%
%[0-1] 
The cluster is named after a tight grouping of four mid-infrared
brightest stars in the cluster
(M1, M2, M3, and M5 in Fig.~1; see also \citealt{Messineo2009}).
We selected the Quartet cluster with the following two criteria.  (1) It
has an $R_g$ that is about half that of the sun, and so a notable
environmental difference compared to the solar neighborhood is
expected. (2) It is important to study clusters
with 
%
%[2] 
more common mass ($\lesssim$$10^3$\,$M_\odot$) for direct comparison
with solar neighborhood ($D \lesssim 2$\,kpc), where most young clusters
have the cluster mass of less than $10^4$\,$M_\odot$
\citep{LadaLada2003}.
However previous studies of clusters in the inner Galaxy had focused on
very massive clusters with $M_{\rm cl} \sim 10^4$\,$M_\odot$, because it
was very difficult to find smaller clusters in such distant regions due
to limited sensitivity.
Recently, however, the Spitzer GLIMPSE survey by e.g.,
\citet{Mercer2005} and \citet{Messineo2009} led to the discovery of
several star clusters of $M_{\rm cl} < 10^4$\,$M_\odot$ such as the
Quartet cluster.

In this paper, we investigated the evolution of protoplanetary disks in
the Quartet cluster.  
In Section~\ref{sec:Quartet}, we describe the Quartet cluster, in
particular, on the age of the cluster, since it is critical for
interpreting the disk fraction.
In Section~\ref{sec:imging}, we present photometry
using moderately deep JHK images obtained from the UKIDSS survey imaging
($K_S \le 17$\,mag).  We then discuss intermediate-mass (IM) stars with
K-band disk excesses (HAeBe stars) discovered in the Quartet cluster in
Section~\ref{sec:results_img}.
For three of the selected HAeBe candidates we obtained IRCS spectroscopy
in Section~\ref{sec:spec}.
We then discuss the properties of objects with IRCS spectra in
Section~\ref{sec:Results_Spec}. Finally, we estimated intermediate-mass
disk fraction (IMDF) for the Quartet cluster to discuss the nature of
disks in the innermost Galactic disk in Section~\ref{sec:IMDF_Qua}.

\section{The Quartet cluster} \label{sec:Quartet}

%\subsection{Previous studies}

\citet{Messineo2009} presented near-infrared (HK-bands) spectroscopic
observations of bright ($K_S \lesssim 10$\,mag) massive stars (Ofpe,
WN9, and OB stars), which are selected based on 2MASS data.  The
distance is estimated to be 6.3\,kpc from OB stars in the cluster.
This is consistent with the kinematic distance derived with radio
recombination line observations to an \ion{H}{2} region G024.83$+$00.10
(6\,kpc; \citealt{Kantharia2007}), with which the Quartet cluster is
associated.  The age of the cluster is estimated to be 3--8\,Myr: older
than 3\,Myr from the existence of evolved stars, such as WR and
Ofpe/WN9, and younger than 8\,Myr from the existence of WR stars and
the absence of red supergiants (RSGs).  They also estimated the cluster
mass to be a few $10^3$\,$M_\odot$ from a simple simulation with
stellar populations.

%\subsection{Notes for the age of the Quartet} 

% \citet{Messineo2009} estimated that the age of the cluster is 3--8\,Myr
%because WRs exist and RSGs do not exist.  

In addition, because both WN and WC stars exist (two WN and one WC), the
age range should be narrowed to $\sim$3--6\,Myr (\citealt{Schaerer1999},
\citealt{Crowther2006}).  Moreover, from the existence of at least three
OB stars whose masses are $\gtrsim$40\,$M_\odot$ (Fig.~4 in
\citealt{Messineo2009}), the age of the Quartet is likely to be less
than 4.5\,Myr because such stars evolve off the main sequence (MS) in
that time \citep{Lejeune2001}.  In summary, the age of the Quartet is
most likely $\sim$3--4.5\,Myr.

\section{IMAGING DATA and SPECTROSCOPY}

\subsection{UKIDSS Survey Imaging} \label{sec:imging}

%\noindent \hspace{-5mm}  {\bf UKIDSS data} 

We used data from the United Kingdom Infrared Telescope Infrared 
Deep Sky Survey (UKIDSS; \citealt{Lawrence2007}) 
Galactic Plane Survey (GPS; \citealt{Lucas2008}) archive, 
which includes J, H, K, and H$_2$-band images.
The UKIDSS images are obtained using the 3.8\,m United Kingdom 
Infrared Telescope (UKIRT), 
and the WFCAM wide-field camera \citep{Casali2007}.
WFCAM pixel scale is 0.4\,arcsec and the images have a FWHM$\sim$1\,arcsec.
We obtained J, H, and K band images around the Quartet cluster 
from the GPS data release 4.
The MKO near-infrared photometric filters (\citealt{Simons2002};
Tokunaga, Simons, \& Vacca 2002) are employed.

%\noindent \hspace{-5mm}  {\bf Photometry}\\

Because the target region is near the Galactic plane and is confusion
limited, the WSA (WFCAM Science Archive) catalogue based on UKIDSS is
found to be incomplete, in particular for fainter stars.
We therefore performed photometry with 
PSF fitting using IRAF/DAOPHOT\footnote{IRAF is distributed by the
National Optical Astronomy Observatories, which are operated by the
Association of Universities for Research in Astronomy, Inc., under
cooperative agreement with the National Science Foundation.}.
%
%
% RC1 [0-2] 
For deriving PSF, we picked up unsaturated bright stars, which are well
off the edge of the frame and do not have any nearby stars with magnitue
difference of more than 4\,mag. PSF photometry was performed using the
ALLSTAR routine two times, once using original images and once using the
subtracted images.  We used PSF fit radii of 3.8, 3.3, and 3.2 pixels
for J-, H-, and K-band, respectively, which are the FWHM values, and set
the sky annulus four times as large as the PSF fit radii.
Saturated stars are removed considering the flag information in the WSA
catalog.  For the photometric calibration, we used about 100 isolated
stars in the frames, whose magnitudes were obtained from the 2MASS Point
Source Catalog, and converted to MKO magnitudes with the photometric
color corrections given by \citet{Leggett2006}.  We also confirmed that
obtained magnitudes are almost identical ($\le$0.02\,mag) to those in
the WSA catalogue for isolated stars.  Limiting magnitudes
(10\,$\sigma$) of $J\simeq 19$\,mag, $H\simeq 18$\,mag and $K \simeq
17$\,mag were obtained.  Using the photometric results here, we
identified eight HAeBe candidates (HAeBe 1--8; red squares in
Fig.~\ref{fig:3col_img} and Tab.~\ref{tab:QUA_HAeBe}), which is
described in detail in Section~\ref{sec:results_img}.

\subsection{IRCS Spectroscopy} \label{sec:spec}

%\noindent \hspace{-5mm}  {\bf IRCS spectroscopy}
For three HAeBe candidates (HAeBe 2, 3, and 5), we conducted
spectroscopic observations using the IRCS (\citealt{Tokunaga1998}, 
\citealt{Kobayashi2000})
%Tokunaga et al. 1998;
attached to the Subaru 8.2\,m telescope on 2012 May 27.
We used the grism mode in combination with the K filter to cover the
wavelength region from 1.93\,$\mu$m to 2.48\,$\mu$m and pixel scale is
52\,mas pixel$^{-1}$.  We used the AO system with natural guide star
mode \citep{Hayano2010} using wave front reference stars, which are
located within 30$''$ from the targets and whose magnitudes are
$R\simeq$14--15\,mag (Fig.~\ref{fig:3col_img}, aqua plus symbols).  We
achieved a FWHM of $\sim$0.2--0.3$''$ in the K band.  We used a slit
with 0.225$''$ width, which provided a spectral resolving power of
$R\sim 800$.

The slit position angles for targets were set to avoid
contaminations of other stars in the slit. The telescope was nodded
along the slit by about 5\,arcsec between exposures.
In order to avoid saturation, exposure time for each frame is set as
180\,sec.
About 10 sets of data were obtained for each target, resulting in a
total exposure time of $\sim$1800\,sec. Spectra of bright telluric
standard stars (HD 139731; F0V) at similar airmass were obtained in 
a similar fashion.
We summarize the details of the observation in
Table~\ref{tab:QUAspec_target}.

%\noindent \hspace{-5mm}  {\bf Data reduction} \\

All the data were reduced following standard procedures using the IRAF
{\tt noao.imred.echelle} package, including sky subtraction (subtraction
of two frames), flat-fielding (halogen lamp with an integrating sphere),
and aperture extraction.  Argon lamp spectra that were taken at the end
of the observing night were used for vacuum wavelength calibration.
Each target spectrum was divided by the spectrum of the F0V standard
star in order to correct for atmospheric absorption and instrumental
response after the Br$\gamma$ absorption line of the standard spectrum
was eliminated with linear interpolation.  
%
% RC1 [10-1] 
Usually, AO correction is more effective for longer wavelength, while
the diffraction-limit FWHM is larger for longer wavelength.  In our
observation, the observed FWHM is found to be almost constant throughout
the band maybe because both effects are cancelled out.
The signal to noise (S/N) for the continuum level for the three targets,
HAeBe 2, 3, and 5 (see Section~\ref{sec:results_img}), are 42, 9, and
46, respectively (Table~\ref{tab:QUAspec_target}).  
Considering the S/N, lines with equivalent width (EW) of $\ge$0.8\,\AA\
is detected with 3\,$\sigma$ for HAeBe 2 and 5, while lines with EW of
$\ge$4\,\AA\ is detected with 3\,$\sigma$ for HAeBe 3.

%\section{RESULTS} \label{sec:results}

\section{HAeBe Candidates in the Quartet cluster}
\label{sec:results_img}

Using photometric results in Section~\ref{sec:imging}, we selected HAeBe
candidates in the Quartet cluster members with the following steps.

%\vspace{1em}
%\noindent \hspace{-5mm}  {\bf Definition of cluster region and control
%field} \\
\subsection{Definition of cluster region and control field}
\label{sec:field}

First, we defined the cluster region and the control field using the
UKIDSS K band image of the Quartet cluster shown in
Fig.~\ref{fig:3col_img}.  We counted the numbers of stars within circles
of a 50 pixel (10$''$) radius around the cluster position.  We then
defined the center of the circle which includes the maximum numbers of
stars as the cluster center, and this was found to be at $\alpha_{\rm
2000} = 18^{\rm h} 36^{\rm m} 17.7^{\rm s}$, $\delta_{\rm 2000} =
-07^\circ 05' 12.1''$, with an uncertainty of $\sim$2$''$.
Fig.~\ref{fig:Pro_Qua} shows the radial variation of the projected
stellar density from the cluster center in the K band.
The horizontal solid line shows the background stellar density.
%
% RC1 [3] 
There is a dip in the plot at the radius of 20$''$.
Such a dip in this kind of radial variation plot is often seen for
brighter stars in nearby clusters (e.g., \citealt{Pandey2008}) probably
due to the stochastic IMF effects on the number of higher mass stars,
thus brighter stars \citep{Santos1997}.
However, in our case, this may be also because the bright stars block
fainter stars in the images so that the number of stars is not correctly
counted.
The very bright stars appear to block fainter stars located within 2.5
FWHM radii ($\sim$2.5\,arcsec) from the centers of the bright stars. The
area of a circle with the radius is about 20\,arcsec$^2$. Because there
are 6 very bright stars (the four brightest stars, M6, and M7 in Fig.~1)
around the center of the cluster, the total area of the blocked region
is 120\,arcsec$^2$ at most.
In the dipped bin in the radial plot ($r = 10''$--20$''$) the total area
is about 950 ($= \pi \cdot (20^2 - 10^2)$) arcsec$^2$ and the blocking
can affect about 13\,\% ($= 120 / 950$) of the total number {\it at
most} and may be partly contributing to the $\sim$20\,\% dip in the
plot.

We define the cluster region to be within a radius of $30''$ where
the stellar density is larger than that of the background region by
3\,$\sigma$.  This corresponds to 0.9\,pc assuming a distance to the
cluster of 6\,kpc.  The defined cluster region is smaller than the
45$''$ radius found by \citet{Messineo2009}, and one of their cluster
members, No.9, is not included in our definition.
However, it is necessary to define the cluster region strictly in order
to obtain cluster parameters with higher S/N by subtracting for a 
control field in the later sections because there should be a lot of
contaminations in the field of the Quartet.
The region at $r\ge 1$\,arcmin is regarded as the background region
with field stars because the stellar density at $r \ge 1$\,arcmin
appears to fall off sufficiently (Fig.~\ref{fig:Pro_Qua}).
We defined a control field at $(\alpha_{\rm 2000}, \delta_{\rm 2000}) =
(18^{\rm h} 36^{\rm m} 14.9^{\rm s}, -07^\circ 06' 32.1''$), which is at
the same Galactic latitude but 1.5\,arcmin off to smaller Galactic
longitude.  
In Fig.~\ref{fig:3col_img} (left), the defined cluster region is shown
with a thick solid circle, while the control field is shown with a
thick dashed circle.

\subsection{Color-magnitude diagram}  \label{sec:col-mag}

We next constructed the $J-H$ versus $J$ color-magnitude diagram of all
detected point sources in the cluster region (Fig.~\ref{fig:CM_Qua},
left) and those in the control field (Fig.~\ref{fig:CM_Qua}, right).
%
% RC1 [4-3] 
The isochrone model by Lejeune \& Schaerer (2001) in the mass range of
$\ge$3\,$M_\odot$ for an age of 4\,Myr and those by Siess et al. 2000 in
the mass range of 0.1--7\,$M_\odot$ for the same age are shown with
magenta and blue thin lines, respectively.
Stars with the mass of $\sim$3\,$M_\odot$ are on the turn-on point to
the main sequence for the age. 
A distance of 6\,kpc was assumed.
An arrow shows the reddening vector of $A_V = 5$\,mag.  In the
color-magnitude diagram the extinction $A_V$ of each star was estimated
from the distance along the reddening vector between its location and
the isochrone models by \citet{Lejeune2001}.
%

%We checked that  
%Stars with mass of more than $\sim$3.5\,Mo are already in MS phase in
% the age of 3--4.5\,Myr, although very massive stars end their lives 
%in the age range, $>$120Mo, 46, 40, for 3, 4, and 4.5\,Myr,
% respectively. 
%
%
%
%%%

We then constructed the distributions of the extinction of stars in the
cluster region (black) and those in the control field (gray) in
Fig.~\ref{fig:Av_Qua}.  The resultant distribution for the control field
shows three peaks at $A_V = 8$\,mag, 14\,mag and $\sim$24--30\,mag,
while that for the cluster region show a significant excess in the $A_V$
range of 14--24\,mag.  
The peak value of the excess extinction for the stars in the cluster
region, $A_V = 16$\,mag, is consistent with \citet{Messineo2009}, who
estimated extinction of the cluster as $A_K = 1.6$\,mag.  
Therefore, we identified the cluster members as being in the cluster
region and with $A_V$ of 14--24\,mag.
However, because there could still be a significant amount
of background contamination from field stars, we correct for this in
discussion in Section~\ref{sec:IMDF_Qua}.

In Fig.~\ref{fig:CM_Qua}, candidate cluster members, stars with $A_V$ of
14--24\,mag, are shown with filled circles, while other stars are shown
with gray squares.
The average value of extinctions for the possible cluster members is
estimated to be 18\,mag.  The isochrone model by \citet{Lejeune2001} for
the age of 4\,Myr with the distance of 6\,kpc and the average extinction
is shown with magenta thick line, while those by \citet{Siess2000} for
the same age with the same distance and extinction is shown with blue
thick line.  The short horizontal lines placed on the thick magenta
lines show positions of
45, 30, 20, 10, 7, 5, and 3\,$M_\odot$, that are also shown on the right
 y-axis.  This indicates that the mass detection limit of possible
 cluster members with limiting magnitude of $J \simeq 19$\,mag is
 $\sim$5\,$M_\odot$.
%
% RC1 [4-1]
Therefore, all the possible cluster members should be in the MS phase at
the age of 4Myr.
Just in case, we checked the isochrone tracks in the age range of
$\sim$3--4.5\,Myr (see Section~2) to see if this conclusion holds.
The isochrone tracks for the ages of 3 and 5Myr are shown with dark gray
lines in Fig.~\ref{fig:CM_Qua}.
%
% RC1 [4-2]
Because isochrone tracks of the PMS phase are located below the
distribution of all the candidate cluster members in the color-magnitude
diagram, all the possible cluster members should be in the MS phase
regardless of age.
Therefore, the estimated extinction values are not
affected by the change of isochrone tracks with age.
%

% RC1 [5-2]
 In Fig.~\ref{fig:CM_Qua}, we also plotted the brightest stars in the
cluster in \citet{Messineo2009} with large orange triangles, filled ones
for possible cluster members and open ones for field stars.
Note that two stars in \citet{Messineo2009}, No. 4 and 9, are not shown
in Fig.~\ref{fig:CM_Qua} because they are not located in the cluster
region with 30$''$ radius, but are at large distances of $\sim$1$'$ and
$\sim$2$'$ from the cluster center, respectively.
%
%\footnote{\color{red} In \citet{Messineo2009}, No. 4 is
%  not identified as a cluster member, while No. 9 is
%  identified as a cluster member. However, No. 9 may be
%  a non-cluster member because it is separated by more
%  than 2$'$ (outside of left figure of Fig.~1), four
%  times of the cluster radius, from the cluster center
%  and because stars as bright as No. 9 are widely
%  distributed (see Fig.~1, left). However, we cannot
%  reject the possibility of the cluster member because
%  \citet{Messineo2009} suggested that it is spectral
%  type is OB type stars, which is usually seen in young
%  clusters.}
%
Only a very bright star, No. 5, which is identified as a cluster member
in \citet{Messineo2009}, appears to have a large extinction of $A_V \sim
30$\,mag (see large orange filled triangle in Fig.~3), which is out of
the extinction range defined above.
Because this star is identified as a late-type WC star in
\citet{Messineo2009}, the large extinction can be attributed to the
thick circumstellar shell.
In fact, the large color excesses of $\Delta (J-H) \sim +0.5$--1.5\,mag 
% and $\Delta H-K \sim 0$--1\,mag 
%for the dust-forming WC stars compared to non-dusty stars 
are observed in case of WC stars in Westerlund 1 cluster with the age of
$\sim$4.5--5Myr \citep{Crowther2006},
which is consistent with the differences between colors of the No. 5 star
and those of other cluster members identified in \citet{Messineo2009},
No. 1, 2, 6, and 7 ($\Delta (J-H) \sim 1$\,mag; see
Fig.~\ref{fig:CM_Qua}).  
Therefore, the apparently large color excess of the No. 5 star is likely
to be from its intrinsic cause.

In Fig.~3, we also plotted sources in 2MASS Point Source Catalog, which 
are in the cluster region and for which the photometry cannot be
performed with UKIDSS data because of saturation, with small orange
triangles.
Only the stars with good photometric quality (the flag of `A' and `B')
are shown.
The results that there are few stars with $A_V > 24$\,mag (2 out of 12)
also support the idea that the $A_V$ range of 14--24\,mag can pick up
almost all of the cluster members.

%\vspace{1em}
%\noindent \hspace{-5mm} {\bf Picking up intermediate-mass stars} 
\subsection{Selection of intermediate-mass stars} \label{sec:IM_select}

Before identifying HAeBe candidate stars (IM stars with K-band disk
excess as defined by \citealt{Hernandez2005}), we selected IM stars
from the candidate cluster members identified in
Section~\ref{sec:col-mag} based on the $J-(J-H)$ color-magnitude diagram
(Fig.~\ref{fig:CM_Qua}).
%
%
% RC1 [9] 
The J-band magnitudes are less affected by disk excess
emission\footnote{Just in case, we confirmed this with a young cluster,
Taurus with the age of 1.5\,Myr. In this cluster, 9 HAeBes and 20
IM-stars without disks in the spectral range of B9--K5 exist
\citep{Yasui2014}. We found the average J-band magnitude of HAeBes to be
9.0$\pm$1.5\,mag, while that of IM stars without disks to be
8.7$\pm$1.0\,mag, suggesting no significant difference of J-band
magnitudes. Considering that the differences of distance modulus of
stars in a cluster are negligible and that the spectral types of HAeBes
and IM stars without disks are not significantly different, the J-band
excess of HAeBes should be negligible.},
while large H and K band excesses are seen in HAeBes (IM stars with
K-band excess; see Fig. 5--9 in \citealt{Hernandez2005}).
Moreover, the effects of differential extinction should be very small
as discussed in Section~4.2. Therefore, we used the J-band magnitudes
 for selecting IM stars as is usually done for studying distant open
 clusters. 
This approach was successfully applied to the studies of old clusters
e.g., $\sigma$ Orionis ($\sim$3\,Myr; \citealt{Hernandez2007},
\citealt{Lodieu2009}), Upper Scorpius ($\sim$5\,Myr;
\citealt{Aller2013}).

The presence of disks around stars earlier than B5
($\simeq$6--7\,$M_\odot$ in the main-sequence phase) is not well
established since the disk lifetime of high-mass stars is probably very
short, e.g. $\sim$1\,Myr \citep{{Zinnecker2007},{Fuente2002}}. 
Therefore, we set the highest target mass of IM stars as 7\,$M_\odot$
that still have protoplanetary disks. 
This is the same upper limit mass as \citet{Yasui2014}, who
systematically derived IMDFs for young clusters in the solar
neighborhood and estimated disk lifetime of IM stars.

Using isochrone models by \citet{Lejeune2001} with the age of 4\,Myr
assuming the distance of 6\,kpc and the extinction of $A_V = 18$\,mag,
the stellar mass of 7\,$M_\odot$ corresponds to $J = 18$\,mag.
Because stars with mass of more than 5\,$M_\odot$ are already in MS
phase, the difference of J-band magnitudes of stars with different age
between 3 and 4.5\,Myr are very small, only $\lesssim$0.05\,mag.
Therefore, we selected faint stars with $J \ge 18$\,mag as IM stars,
which are shown with large filled circles in Fig.~\ref{fig:CM_Qua}. The
limiting magnitude of UKIDSS images is about $J=19$\,mag, which
corresponds to about 5\,$M_\odot$.

%\vspace{1em}
%\noindent \hspace{-5mm} {\bf Color-color diagram}  
\subsection{Color-color diagram} \label{sec:cc}

% RC1 [12] 
K-band excess of YSOs suggests the existence of innermost dust disk
($\sim$1\,AU in the case of Herbig Be stars;
\citealt{Millan-Gabet2007}), where the dust is in the sublimation
temperature ($\sim$1500 K). 
The $J-H$ versus $H-K$ color-color diagrams for the stars in the cluster
region and in the control field are shown in Fig.~\ref{fig:CC_Qua}.
Stars with the extinction of $A_V = 14$--24\,mag (probable cluster
members) and the other stars (mainly field stars) are shown with filled
circles and small gray squares, respectively.  Among the likely cluster
members, stars with $J\ge 18$\,mag (IM stars) are shown with large
filled circles.
The line that intersects $(J-H, H-K) = (0.2, 0)$ and is parallel to the
reddening vector,
%\textcolor{blue}{(gray dot dashed lines in Fig.~\ref{fig:CC_Qua})}, 
is the borderline between IM stars with circumstellar disks (candidate
HAeBe stars; right side) and without circumstellar disks (left side)
\citep{Yasui2014}.
This line is defined to avoid selecting the reddest star
without disk among all the samples in \citet{Hernandez2005}, HIP
112148. 
%alghouth this line cannot necessarily
%select all of stars with disks (see Fig.~2 in \citealt{Hernandez2005}).} 
%
However, because this borderline is originally defined in the 2MASS
system, we checked the location of the line in the MKO system, which is
used by UKIDSS.
We converted the 2MASS magnitudes of the HAeBe stars and the classical
Be stars for all samples in \citet{Hernandez2005} to MKO magnitudes with
the color transformation equations given by \citet{Leggett2006}.
Considering the converted colors of HIP 112148, the MKO borderline is
defined as the line that intersects $(J-H, H-K) = (0.22, 0)$ and is
parallel to the reddening vector (dot dashed lines in
Fig.~\ref{fig:CC_Qua}).
%
%We also confirmed that stars without disks and stars with disks are
%located at left and right side of this line, respectively. 
%
Because very few stars with disks in \citet{Hernandez2005} are bluer
than the borderline (e.g., the filled circle with $(H-K)_0$ of
$\sim$0.2\,mag in Fig.~5 by \citealt{Hernandez2005}), the number of
selected HAeBe candidates with this borderline should be lower limit.
Indeed, some groups e.g., \citet{Comeron2008} defined the bluer borderline
(the grey dot-dashed line in Fig.~1 in \citealt{Yasui2014}), which
can select more HAeBe candidates. 
Using the MKO borderline, we selected eight stars with possible K disk
excesses among the probable IM stars in the cluster region.
These are the candidate HAeBe stars in the Quartet cluster, and they are
shown with red filled circles in Figs.~\ref{fig:CM_Qua} and
\ref{fig:CC_Qua}.
The coordinates and magnitudes of the HAeBe candidates are summarized in
Table~\ref{tab:QUA_HAeBe}, and their positions are shown with red
squares in Fig.~\ref{fig:3col_img} (right).

\section{Spectroscopy of three HAeBe candidates} \label{sec:Results_Spec}

We obtained K-band spectra of HAeBe 2, 3, and 5 as described in
Section~\ref{sec:spec}.  The location of these three stars in the
color-color diagram is shown with filled circles enclosed by open
circles in Figs.~\ref{fig:CM_Qua} and \ref{fig:CC_Qua}. They are
relatively bright in the K-band and are located close to bright stars
($R\simeq 14$--15\,mag) that are needed as natural guide stars
for AO mode observation.
The spectra of three HAeBe candidates, which were normalized to 1.0
after 2-pixel smoothing, are shown with an offset for clarity in
Fig.~\ref{fig:Spec_Qua}. 
In all spectra, the Br$\gamma$ absorption lines at 2.166\,$\mu$m and
first overtone CO bandhead emissions at 2.294\,$\mu$m are detected.
EWs of the Br$\gamma$ absorption lines were measured between 2.160 and
2.175\,$\mu$m, while those of the CO bandhead emission lines were
measured between 2.290 and 2.315\,$\mu$m. 
The uncertainties are estimated as $\Delta \lambda N_{\rm pix}^{1/2}
({\rm S/N})^{-1}$, where $\Delta \lambda$ is the spectral dispersion of
%6.13729899999971
6.14\,\AA, $N_{\rm pix}$ is the number of pixels used for measuring
EWs, and S/N is the pixel-to-pixel S/N of the continuum.
The estimated EWs and their uncertainties of the three spectra are
summarized in Table~\ref{tab:QUAspec_target}.}
%
%The uncertainties of the EWs are estimated using IRAF splot task with
% the spectral dispersion, S/N, and the number of pixels occupied by
% features \citep{Kulkarni2005}. 
%del_lam = 2um / 800 * 10^4 = 25
%
%
%
%
%
% RC1 [10-1] 
 We also show the obtained spectrum of the telluric standard star, HD
139731, in the bottom plot in Fig.~\ref{fig:Spec_Qua}.
The observed spectrum, which was reduced in the same way
as in Section~\ref{sec:spec} but was not divided by
spectrum of standard star, is shown with black, while
that after eliminating the Br$\gamma$ absorption line
(see Section~\ref{sec:spec}) is shown with gray.
The pixel-to-pixel S/N for the continuum level of the
spectrum is estimated to be as high as 167.
%The estimated EW of the Br$\gamma$ absorption line is 7.6$\pm$0.2.
%The EW of Br$\gamma$ is estimated to be 7.6$\pm$0.2. 
%Hanson1996よりmid B or later, Ali1995よりlate B to late F. 

\subsection{Implication to spectral types} \label{sec:Results_Spec_SpT}

% RC1 [10-4] 
We estimated spectral types of HAeBe 2, 3, and 5 from the obtained
spectral features by comparing them with those in the literatures,
\citet{Hanson1996} for O and B type stars, \citealt{Ali1995} for late B
to M type stars, and \citet{Rayner2009} from F to M type stars.
%
% RC1 [10-3] 
The spectral resolutions in the references are
comparable or higher ($R\sim$ 800--3,000) compared to
our data ($R\sim$ 800).

% RC1 [10-4] 
All three HAeBe candidates show Br$\gamma$ absorption line, which are
 known to appear clearly in stars from O to G type (\citealt{Hanson1996}
 for O and B type; \citealt{Ali1995} for late B to G type).  The EW
 reaches a maximum at a spectral type of
%late B to A (${\rm EW} \sim 10$\,\AA\ from \citealt{Hanson1996}; ${\rm
%EW} \sim 8$\,\AA\ from \citealt{Ali1995}) 
A0--A5 type (${\rm EW} \sim 10$\,\AA\ by \citealt{Hanson1996}; ${\rm EW}
\sim 8$\, by \citealt{Ali1995}), and then becomes smaller with later
type. 
%
% RC1 [10-4] 
For early type stars in this spectral range, O type stars are known to
show \ion{N}{3} emission lines at 2.116\,$\mu$m and \ion{He}{2} at
2.188\,$\mu$m \citep{{Hanson1996},{Hanson2005}}, while B type stars show
\ion{He}{1} at 2.113\,$\mu$m in stars earlier than B3
\citep{Hanson1996}.
The non-detection of the above lines with ${\rm EW}
\gtrsim 0.8$\,\AA\ for HAeBe 2 and 5 (see
Section~\ref{sec:spec}) suggest that they are late O
type or later, although the detection of the above two
lines
%other absorption lines than Br$\gamma$ 
is difficult for HAeBe 3 due to the low S/N of $\sim$10.
On the other hand, F or later type stars are known to
show absorption lines of \ion{Ca}{1} at 2.263 and
2.266\,$\mu$m, \ion{Na}{1} at 2.206 and 2.209\,$\mu$m,
and \ion{Mg}{1} at 2.279--2.285\,$\mu$m
%which are characteristics of F type and later type stars
\citep{{Ali1995},{Rayner2009}}.  
The features become more prominent in later type stars
to be comparably strong as the Br$\gamma$ absorption
feature in K type stars.
% . Around in K
Because all the three HAeBe candidates do not show
\ion{Ca}{1}, \ion{Na}{1}, and \ion{Mg}{1}, the spectral
types of the three are most likely to be G or earlier.
In summary, the suggested spectral types for HAeBe 2 and
5 are late O to G, while that for HAeBe 3 is O to G.

% RC1 [10-2] 
 We also estimated the spectral types of HAeBe 2, 3, and 5 from the EWs
of Br$\gamma$ absorption line.
Because the EWs of Br$\gamma$ by \citet{Ali1995} was measured in
2.160--2.170\,$\mu$m, which does not necessarily cover the Br$\gamma$
lines, we also measured the EWs in the same wavelength range for HAeBe
2, 3, and 5 as 3.6$\pm$0.6, 7.7$\pm$2.8, 5.4$\pm$0.6\,\AA, respectively,
and then compared to the EWs with those by \citet{Ali1995} for
estimating spectral types.
%, spectral types are
%estimted to be G for HAeBe 2, late B to early G for HAeBe 3, and late B
%or late F to G for HAeBe 5.\\
%
Although we could not find the wavelength range used for measuring EWs
by \citet{Hanson1996}, the larger EWs by \citet{Hanson1996} than those
by \citet{Ali1995} for the same spectral types, $\sim$2\,\AA, may be due
to the wider range by \citet{Hanson1996}.
Therefore, we compared EWs originally measured between 2.160 and
2.175\,$\mu$m (Table~\ref{tab:QUAspec_target}) with those by
\citet{Hanson1996}. 
%
%
% RC1 [10-2] 
First, as a consistency check, we measured the EWs for observed standard
stars with known spectral type.  Besides the telluric standard HD 139731
(F0V), we obtained a spectrum of another star, G155-23 (K7V) with high
S/N. We show their spectra, which were reduced in the same way as in
Section~3.2 but was not corrected for the telluric absorption, in the
bottom plot of Fig.~\ref{fig:Spec_Qua}.
For HD 139731, the estimated EW of Br$\gamma$ is 6.9$\pm$0.1 in
2.16--2.17$\,\mu$m and 7.6$\pm$0.2 in 2.16--2.175um, which is consistent
with the F0 type stars (EW of 6.5$\pm$1 in \citealt{Ali1995}).
%although the estimated spectral type from the EW has very wide range
%considering errors, late B to late F. 
Also, the non-detection of Br$\gamma$ line in G155-23 is
consistent with its spectral type (\citealt{Ali1995},
\citealt{Rayner2009}).
We therefore estimated spectral types for HAeBe 2, 3, and 5 to be G or O
to early B, B to G, and late O to early B or late F to G, respectively.
However, it should be noted that objects with protoplanetary disks often
have hydrogen emission lines due to mass accretion activity
\citep{Donehew2011} and thus the EWs of the central stars could be
larger than the observed EWs for low-resolution spectroscopy.
Therefore, the true spectral types of the above stars could be closer
to A0--A5.

In summary, detections and non-detections of absorption lines suggest
that the spectral types for all three HAeBe candidates are late O to G,
though this is still a rather wide range for spectral type due to the
limits of the currently available data.
In the age of 3--4.5\,Myr, the spectral type of G corresponds to the
stellar mass of $\gtrsim$2.0\,$M_\odot$
\citep{Siess2000},
which is consistent with the intermediate-mass we aimed only using the
J-band magnitudes in Section~\ref{sec:results_img}
($\ge$1.5\,$M_\odot$). 
Although we could not put strong constraints on the upper limit of the
stellar mass by spectroscopy (late O type corresponds to
$\sim$20\,$M_\odot$), the three objects should not be
background/foreground massive stars in view of the very small
probability of chance projection.
In fact, the J-band magnitude of late O type stars is $\simeq$15.5\,mag
at the distance of 6\,kpc and the extinction of $A_V = 18$\,mag, which
is much brighter than the magnitudes for IM-stars ($J\sim 18$\,mag for
7\,$M_\odot$).
Because the magnitude for stars with mass of 10\,$M_\odot$ comes down to
$J\sim 17.4$\,mag, which is only $\sim$0.5\,mag brighter than the
maximum magnitude of IM stars, the upper limit mass of the three objects
are likely to be $\sim$10\,$M_\odot$ at most, corresponding to early B 
type.

% 20Mo: J=15.5
% 10Mo: J=17.3787562519182 

% MJ=-7.2 (@20Mo), DM=13, Aj = 0.282 * 18 mJ=10.9
% MJ~-5 (@10Mo), DM=13, Aj = 0.282 * 18 mJ=10.9

%\subsection{Nature of three HAeBe candidates with spectra}
\subsection{Nature of HAeBe 2, 3, and 5}
\label{sec:prop_spec} 

All three HAeBe candidates were found to have prominent CO bandheads in
{\it emission}.  The K-band excess noted in Section~\ref{sec:cc} might be due
to the CO bandhead emission and so we investigated this possibility.
The spectra of HAeBe 2, 3, and 5 were divided by the spectrum of a F0V
standard star and multiplied by a blackbody function of 7,300\,K
\citep{Drilling2000} to obtain the relative flux density.  
We then calculated the ratio of the flux from the CO bandhead emission
to the continuum flux in the wavelength range of 2.03--2.37\,$\mu$m
where atmospheric transmission is
$\ge$50\,\%\footnote{http://www.ukidss.org/technical/technical.html}.  
As a result, the ratios were found to be 5\,\%, 3\,\%, and 2\,\% in
HAeBe 2, 3, and 5 respectively.  If we subtract the excess CO bandhead 
emissions, all of the objects still have sufficient K-band excesses to
be identified as HAeBe candidates on the JHK color-color diagram.

The CO bandhead emission is characteristic of a stellar object harboring
dense circumstellar material.  The conditions required to excite the
overtone emission are high temperatures ($\sim$2,000\,K) and high
densities ($\gtrsim$$10^{10}$\,cm$^{-3}$; \citealt{Scoville1980}).  CO
overtone emission is detected in young stellar objects (YSOs;
\citealt{{Scoville1979},{Scoville1983}}), supergiants (SGs;
\citealt{McGregor1988}), and proto-planetary nebulae (PPNe;
\citealt{Hrivnak1994}).
However, the positions of SGs in the $J-(J-H)$ color-magnitude diagram
(Section~4.3) are very different from those of IM stars ($\Delta J \sim
7$\,mag; see Fig~3 and Fig.~1 in \citealt{Bik2006}).
Also, SGs with K-band excess (B[e]SGs) are known to show
Br$\gamma$ {\it emission}, which is not observed for
HAeBe 2, 3, and 5.
As for PPN, it should not be associated with the Quartet
cluster because the age of the central star of PPN is
measured in Gyr ($>$10\,Gyr for the case of
1\,$M_\odot$; \citealt{Schroeder2008}). Moreover,
because the period of the SGs and PPNe phase are very
short ($\sim$$10^4$\,yr for SGs by \citealt{Smartt2009};
$\sim$$10^3$\,yr for PPNe by \citealt{Kwok1993}), it is
highly unlikely that such rare objects in the foreground
or background are coincidentally superposed on the
Quartet cluster region.
On the other hand, the Quartet cluster is a young star
cluster and the properties of HAeBe 2, 3, and 5 are
consistent with YSOs from photometry
(Section~\ref{sec:results_img}) and spectroscopy
(Section~\ref{sec:Results_Spec_SpT}).
Therefore, HAeBe 2, 3, and 5 are most likely to be YSOs in the Quartet
cluster.

The fraction of YSOs with CO emission is quite low \citep{Najita2007}. 
Although we could not find any good compilations of K-band spectra for
YSOs, it would be useful to compare our results to several references.
\citet{Hoffmeister2006} estimated 4\,\% for M17, which is a very young
star-forming cluster with an age of 1\,Myr, although the mass range of their
samples is on the higher side ($\sim$5--20\,$M_\odot$) compared to the
Quartet cluster ($\sim$5--7\,$M_\odot$).
\citet{Ishii2001}, whose targets are IRAS sources identified as luminous
YSOs, estimated fraction of the detection at 15\,\% for HAeBe stars.
In the case of the Quartet cluster, the fraction is at least
$\sim$40\,\% considering that there are eight HAeBe candidates and all
three objects with spectra has CO emission (38\,\% (3/8)).
This value appears to be quite high and this may be characteristic of
IM stars in the inner Galaxy.
However, because the possibility of variability in both CO emission and
Br$\gamma$ lines is suggested \citep{Hoffmeister2006}, further study is
necessary to have a firm conclusion. 
In any case, the existence of both K-band excess and CO bandhead
emission in HAeBe 2, 3, and 5 suggest that they
still have significant amount of dust and gas in their disks.

Br$\gamma$ emission is a well-known tracer of mass accretion
\citep{Muzerolle1998} and indicates the existence of gas disks very
close to the central stars ($<$0.1\,AU; \citealt{Hartmann2009}).
However, the line is less sensitive to mass accretion,
$\simeq$$10^{-8}$\,$M_\odot$\,yr$^{-1}$ or higher
\citep[e.g.,][]{{Muzerolle1998},{Calvet2004}}, compared to optical
hydrogen lines such as H$\alpha$,
$\simeq$$10^{-11}$\,$M_\odot$\,yr$^{-1}$ or higher
\citep{Muzerolle2005}.  Since we could not detect any Br$\gamma$
emission lines, whether the innermost gas disks exist or not for HAeBe
2, 3, and 5 is not known at this stage.

\section{Discussion}

\subsection{Implication to the age of the Quartet cluster} 

 In general, young clusters have high intra-cluster extinctions with
large dispersions
%whereas clusters in the age of 3--4.5\,Myr do not have any considerable
%extinctions 
on a timescale of embedded phase (about 2 to 3\,Myrs;
\citealt{LadaLada2003}):
{$A_V$ from 0 mag to $\sim$25--100\,mag for the age of 1\,Myr, $A_V$
from 0 mag to 5--10mag for the age of 2--3Myr, and $A_V$ from 0 mag to
$<$2--4\,mag for the age of $\ge$4\,Myr}\footnote{We checked extinctions
for all young clusters listed in \citet{Yasui2014}, with reviews of each
star-forming region in \citet{{Reipurth2008hsf1},{Reipurth2008hsf2}} and
references listed in Tab.~3 of \citet{Yasui2014}.
Only one cluster, sigma Ori (3\,Myr old), has very low extinction for
the age of the cluster,
%0.04 mag ＜ E(B − V) ＜ 0.09 mag, 
with maximum $A_V$ of 0.3\,mag. 
Because of small distances to nearby clusters, $\sim$400\,pc on average,
extinctions should be dominated by intra-cluster extinctions.}.
%: Av < 0.32? 
%low extinction (0.04 mag < E(B − V) < 0.09 mag – Béjar et al. 2004;
%Sherry et al. 2008); AV=3.2E(B-V)
%\citealt{Sicilia-Aguilar2005}). 
%and Orion OB1bc (4Myr,\citealt{Hernandez2005}).
%
%
% 
Considering the extinction dispersion of the Quartet cluster, $\Delta
A_V \sim \pm 5$\,mag (Section~\ref{sec:col-mag}), the cluster may be
still in embedded phase.
From $^{13}$CO data by Boston University--Five College Radio Astronomy
Observatory (BU-FCRAO) Galactic Ring Survey \citep{Jackson2006},
% with sensitivity of <0.4 K, spectral resolution of 0.2 km/s, and
% sampling of 22",  
five molecular components are detected in the direction of the Quartet
cluster with peak velocities of $v_{\rm LSR} = 7$, 45, 52, 100, 107\,km
s$^{-1}$.
Considering the location of the Quartet cluster ($l \sim 25^\circ$ and
$D=6$\,kpc), the velocity is likely to be $\sim$100--120\,km s$^{-1}$ in 
the case that the cluster is in nearly circular Galactic orbit,
while it can be $>$70\,km s$^{-1}$ in the case that the cluster has a
large peculiar motion \citep{Reid2014}. 
Although we do not know the velocity of the cluster, even the strongest
component at $v_{\rm LSR} > 70$\,km s$^{-1}$, one with peak $v_{\rm
LSR}$ of 100 \,km s$^{-1}$, has an integrated intensity of 5.4\,K km
s$^{-1}$, corresponding to only $A_V = 3.3$\,mag \citep{Pineda2008}.
Therefore, the molecular component can cause the intra-cluster 
extinction of only $A_V \sim 0$--3\,mag at most even if the cluster is
located in the molecular component.
Compared to the intra-cluster extinction dispersions of young nearby
clusters, the age of the cluster is suggested to be $\gtrsim$4\,Myr, at 
least. 
Distant clusters can also have large extinction dispersions only due to
foreground components, clouds of atomic gas and/or molecular clouds:
%by a large cloud of atomic gas rather than a relatively small dense
%extinction.
% even if they already ended embedded phase. 
% (, Figer+2006)
e.g. $\Delta A_V \sim \pm 4$\,mag for RSGC 1
\citep{Figer2006}\footnote{RSGC 1 is a $\sim$7--12\,Myr old cluster with
identical distance ($\sim$6\,kpc) and the size ($\sim$1.5\,arcmin
radius) to the Quartet. It is located within 30\,arcmin of the Quartet
on the sky, $(l, b) = (25.27, -0.2)$.
The extinction dispersion is estimated from $\Delta (H-K) \sim \pm 
0.25$\,mag in Fig.6 of \citet{Figer2006}. 
In the direction of the cluster, no molecular clouds are detected in the
BU-FCRAO $^{13}$CO survey in the velocity of $v_{\rm LSR} > 70$\,km
s$^{-1}$ with 3$\sigma$ upper limit of 0.082\,K km s$^{-1}$,
% (0.13\,K * 3 * 0.21 (delta v)) K km s-1
corresponding to $A_V = 0.16$\,mag.}.
Therefore, the large extinction dispersion for the Quartet cluster can
be attributed to foreground extinction.

{\{2\}}
We also compare plots of the possible cluster members with the isochrone
tracks in the wide age range of 1--10\,Myr on J-H vs. J color-magnitude
diagram (Fig.~\ref{fig:CM_Qua}).
The isochrone tracks for the ages of 1, 2, 3, 5, 7, and 10\,Myr are
shown with dark and light gray lines in Fig.~\ref{fig:CM_Qua} (left).
The inset shows a blowup of the region enclosed with the black box. The
gray italic numbers in the inset show the ages of isochrone models.
The maximum mass of isochrone models by \citet{Lejeune2001} are set as
120\,$M_\odot$ for the age of 1 and 2\,Myr. 
For these ages, the J magnitudes of the maximum mass stars
($J\sim12$\,mag) are much fainter than those of No. 1 and 2 stars, whose
masses are estimated at $\sim$25--60\,$M_\odot$ from the spectral types
\citep{Messineo2009}, thus the cluster can never be younger than 2\,Myr.
In the contrast, for the age of $\ge$3\,Myr, the maximum masses of the
isochrone models are set as 105, 35, 25, and 18\,$M_\odot$ for the age
of 3, 5, 7, and 10\,Myr, respectively, which are determined by the
lifetime of massive stars.
For these ages, isochrone models cover the magnitude range of all
members of the Quartet cluster, suggesting that the age of the cluster
is $\ge$3\,Myr.
%
%
%Considering suggested initial mass of No. 1 and 2 stars from the
%spectral types, $\sim$25--60\,Mo (Messineo+2009), such younger age seems
%to be unlikely.
%
%The descrepancy of the magnitudes are not explained with distance
%uncertainty of 2kpc (Messineo+2009), corresponding to distance modulus
%difference of 0.7\,mag.
%
Although the isochrone track for the age of 5\,Myr appears to match well
the plots for No. 1, 2, and 6 stars, we cannot put strong constraints on
the age considering the relatively large $A_V$ dispersion
($\sim$5\,mag). 
In any case, the suggested age here ($\ge$3\,Myr) is consistent with the
age estimate in Section~2. 
%
%For the age of $\ge$3\,Myr, 
%However, because there are $A_V$ dispersion of the possible cluster
%members, we cannot make strong constraint on the older age from the
%diagram.  

In summary, most likely age of the Quartet cluster is $\ge$3\,Myr,
supporting the previous age estimates in Section~2 based on WR stars
(3--4.5\,Myr). 

%, although we cannot put any constraints on the older ages.

\subsection{Disk fraction in the Quartet cluster} 
\label{sec:IMDF_Qua}

The fraction of sources having protoplanetary disks
within a young cluster, the disk fraction, is one of the 
fundamental parameters characterizing the evolution of disks and planet
formation. 
We derived the disk fraction of IM stars (IMDF) in the mass range of
$\sim$5--7\,$M_\odot$ for the Quartet cluster using the J-H vs. H-K
color-color diagram (Fig.~\ref{fig:CC_Qua}, see detail in Section~4).
In Section~\ref{sec:results_img} we found eight disk excess sources out
of 42 stars, to yield an IMDF of 19\,\% $(8/42)$. 
Because there could be a significant number of contaminating 
IM field stars in the direction of the Quartet cluster, 
we estimated disk fraction of the Quartet cluster by subtracting 
the estimated number of IM field stars in a control field.
The IMDF in the control field was found to be 0\,\% out of 6 IM stars by
using the same selection methods as for the cluster region. 
As a result, IMDF for the Quartet is estimated to be $22 \pm 8$\,\%
$(8/(42-6))$. 
Just in case, we checked another control field, which has the same area
as the cluster region and is separated by 1\,arcmin to the north (the
thin blue dashed circle in Fig.~1).  For the field, the IMDF was found
to be 0\,\% out of 18 IM stars, and the resulting IMDF for the Quartet
cluster is $33 \pm 12$\,\% $(8/(42-18))$.  This suggests that the IMDF
of $\sim$25\,\% is a reasonable estimate for the Quartet cluster.

In Section 4.1, we discussed that a part of the cluster members could be
missed by the blocking of very bright stars. If the maximum density of
the cluster ($\sim$200 stars per arcmin$^{-2}$ from Fig.~2) is assumed,
at most 7 cluster members could be missed from our sample. Even if all
of those missed members do not have K-band excess, the IMDF is estimated
to be 19\,\% $(8/(42-6+6))$, which is well within the statistical
uncertainty of the originally estimated IMDF.
Another possible source of systematic uncertainty is the definition of
the cluster region. Because the centroid of the cluster may be different
due to the possible blocking of the cluster members, just in case, we
additionally estimated IMDF with an alternatively defined cluster
region, whose center is at the centroid of very bright cluster members,
No. 1, 2, 5, 6, and 7 in \cite{Messineo2009} (18:36:16.95, -07:05.07.3),
and whose radius is the same as the original cluster region (see the
green dashed circle in Fig.~1). We derived the IMDF in this alternative
cluster region to be 18\,\% $(7/39)$, and the estimated IMDF after the
field-star correction is 21\,\% $(7/(39-6))$, which is identical to the
originally estimated IMDF.
%Therefore, the effects of
%the possible blocked stars on the derived IMDF is not
%significant.

% RC1 [12] 
In Section~\ref{sec:IM_select}, we selected IM stars
simply based on J-band magnitudes ($18 \le J \lesssim
19$\,mag).
% because we confirmed
%that the effects of differential extinction should be very small in the
%age of $\le$3\,Myr and also for stars in large distances of 6\,kpc.
This way of selecting IM stars enabled us to include
HAeBe stars, which have large J-H excesses
%($\Delta J-H \sim 0.5$--1\,mag, \citealt{Hernandez2005}), 
from the sample of IM stars (see
Section~\ref{sec:prop_spec}).
%
%Higher mass stars without disks with large extinction could intrude into
%the IM star region of our definition, or IM-stars with small extinction
%could be missed from the region.
% 
However, just in case, we tried to select IM stars along the extinction
vector to see the difference in the estimated IMDF.  In this case, stars
located at the lower left of the thin dashed lines in
Fig.~\ref{fig:CM_Qua}, which pass through the isochrone track by
\citet{Lejeune2001} at the mass of 7\,$M_\odot$ and are parallel to the
reddening vector, can be regarded as IM stars. Out of our original
sample of IM stars without disks (large filled black circles in
Fig.~\ref{fig:CM_Qua}), stars located at the upper right of the line are
excluded, while stars with $J \le 18$\,mag that are located at the lower
left of the line are included as new samples.
Because the numbers of the selected IM stars are
decreased by eight and two in the cluster region and in
the control field, respectively,
IMDF with the new sample becomes 27$\pm$9\,\% $(8/ ((42
- 8) - (6 -2)))$, which is again consistent with the
IMDF based on the selection criteria of
Section~\ref{sec:IM_select} (22$\pm$8\,\%).

In the IMDF-age diagram in Fig.~\ref{fig:IMDF_Qua}, we show the derived
IMDF of the Quartet cluster (red filled square), as well as of the IMDF
for young clusters in the solar neighborhood (black filled circles),
which are from Fig.~5 (left; red symbols) of \citet{Yasui2014}.
The IMDFs in the solar neighborhood were derived from the near-infrared
color-color diagram for stars that were identified as IM stars 
(1.5--7\,$M_\odot$) by spectroscopy.
The IMDFs derived with only NIR JHK-band are generally low, $\sim$30\,\%
at most for cluster ages of $\le$2\,Myr and decrease down to nearly zero
at $\sim$5\,Myr as shown with a black line in Fig.~\ref{fig:IMDF_Qua}. 
The IMDFs derived from mid-infrared data have very high values of
$\sim$100\,\% for ages of $\lesssim$2\,Myr \citep{Yasui2014}.
The disk fraction for OBA type stars observed in the L-band also have a
relatively high value of $\sim$50\,\% in young clusters
\citep{Stolte2010} although the stellar mass range is on a little higher
side than that in this paper.
Fig.~\ref{fig:IMDF_Qua} suggests that the IMDF of the Quartet cluster is 
one of the highest among all young clusters.
For the age of the Quartet cluster, 3--4.5\,Myr old, its IMDF appears to
be higher than those in the solar neighborhood. 
Note that the stellar mass range for the estimation of IMDF for the
Quartet is on the higher-mass side ($\sim$5--7\,$M_\odot$) than those in
the solar neighborhood (1.5--7\,$M_\odot$).
Because shorter disk lifetime is suggested for higher mass stars
\citep{Williams2011}, the estimated IMDF of the Quartet cluster could be
rather underestimated.

Even considering the above uncertainties, the IMDF of the Quartet is
suggested to be one of the highest among IMDFs of young clusters in the
solar neighborhood.
This might be one of the characteristics of clusters in the inner
Galaxy, but should be investigated by observations of more clusters in
the inner Galaxy with a wide range of ages in the future.

\section{Conclusion}

We aimed to investigate environmental effects on the evolution of
protoplanetary disks.
First, we used UKIDSS survey data to search for HAeBe stars in the
Quartet cluster in the innermost Galactic disk at the Galactrocentric
radius of $R_g \sim 4$\,kpc.
We then performed IRCS spectroscopy for some of HAeBe candidates to
confirm their characteristics.
The Quartet cluster is an appropriate target for measuring environmental
effects because the mass of the cluster is more typical
($\sim$$10^3$\,$M_\odot$) compared to previous studies for clusters in
the inner Galaxy with cluster mass of $>$$10^4$\,$M_\odot$.
Although the age of the Quartet is previously estimated to be 3--8\,Myr
old, we find that it is most likely $\sim$3--4.5\,Myr old, because of
the existence of both WC and WN stars and existence of OB stars with
mass of $\sim$40\,$M_\odot$.
Our conclusions can be summarized as follows:

\begin{itemize}

\item As a result of photometry with UKIDSS data with the mass
      detection limit of $\sim$5\,$M_\odot$, 
      we found eight HAeBe candidates with K-band disk excesses in the
      mass range of $\sim$5--7\,$M_\odot$.
      %and have K-band disk excesses, in the
      %Quartet cluster region.

\item For three HAeBe candidates, we performed K-band
      medium-resolution ($R \sim 800$) spectroscopy and found that
      they all have both Br$\gamma$ absorption and CO bandhead
      emission.
      From the existence of Br$\gamma$ absorption line and the
      non-detection other absorption lines, all three are suggested to
      be IM or higher mass stars. 
      The CO bandhead emission suggests that all three objects are most
      likely to be YSOs with protoplanetary disks.
      It should be also noted that the high detection rate of the CO
      emission (at least $\sim$40\,\%), which is generally low in the 
      solar neighborhood ($\sim$15\,\%), may be characteristics of
      disks in such a region.

 \item We estimated IMDF of the Quartet cluster to be $\sim$25\,\%. 
       Although there may be still a large systematic uncertainty, it is
       one of the highest value among IMDF derived for young clusters in
       the solar neighborhood.  Considering the mass detection limit of
       $\sim$5\,$M_\odot$, this value may be even a lower limit.

\end{itemize}

To understand the characteristics of disk evolution in the inner Galaxy,
observation for more young clusters with a wide range of ages as well as
identification of cluster members with less uncertainty by comprehensive
spectroscopic follow-up is important in the future.

\begin{table*}[!h]
%\begin{table}[!h]
%
\caption{HAeBe candidates in the Quartet cluster.}  
\label{tab:QUA_HAeBe}
\begin{center}
\begin{tabular}{llllllllllll}
\hline
\hline
\multicolumn{2}{c}{Star} & R.A. & Decl. & $J$ & $H$ & $K$ \\
& & (J2000.0) & (J2000.0) & (mag) & (mag) & (mag) \\
\hline
HAeBe & 1 & 18:36:18.20 & -07:05:00.2 & 18.05$\pm$0.05 & 16.08$\pm$0.06 
& 14.59$\pm$0.03 \\
& 2 & 18:36:18.18 & -07:04:45.9 & 18.71$\pm$0.09 & 16.42$\pm$0.06 & 
14.72$\pm$0.03 \\
& 3 & 18:36:16.15 & -07:04:59.9 & 18.49$\pm$0.06 & 16.47$\pm$0.04 & 
14.77$\pm$0.03\\
& 4 & 18:36:17.47 & -07:05:15.1 & 18.83$\pm$0.07 & 16.46$\pm$0.06 & 
14.83$\pm$0.02 \\
& 5 & 18:36:16.25 & -07:05:33.0 & 18.44$\pm$0.06 & 16.33$\pm$0.05 &
14.84$\pm$0.05 \\
& 6 & 18:36:17.65 & -07:05:19.1 & 19.10$\pm$0.08 & 16.73$\pm$0.05 &
15.07$\pm$0.02 \\
& 7 & 18:36:17.01 & -07:04:54.3 & 18.56$\pm$0.08 & 16.83$\pm$0.07 &
15.54$\pm$0.05 \\
& 8 & 18:36:18.98 & -07:05:00.9 & 18.84$\pm$0.07 &  17.28$\pm$0.05 &
16.10$\pm$0.05 \\
\hline
\end{tabular}
\end{center}
%{Notes...\\
%De03: DeWarf+2003 }
%
\end{table*}

\begin{table*}[!h]
%\begin{table}[!h]
%
\caption{Spectroscopy log and equivalent width of spectral features.}
\label{tab:QUAspec_target}
\begin{center}
\begin{tabular}{rlclclc}
\hline
\hline
\multicolumn{2}{c}{Target} & Airmass & Integration & S/N$^{\rm a}$ & 
\multicolumn{1}{c}{Br$\gamma$} & CO (2--0) \\
& &  & \multicolumn{1}{c}{(sec)} & & 
\multicolumn{1}{c}{EW (\AA)} & \multicolumn{1}{c}{EW (\AA)} \\
%\multicolumn{1}{c}{$W_\lambda$ (\AA)} & 
%\multicolumn{1}{c}{$W_\lambda$ (\AA)} \\
% & & & (mag) & (mag) & (mag) \\
\hline
% 
% ID7---H2 
HAeBe & 2 & 1.2 & $180 \times 12$ & 42  & 4.4$\pm$0.7 & $-$22.7$\pm$0.9 \\
% &  & &  &  & (3.6$\pm$0.6) & \\
%
%
% ID6---H3
& 3 & 1.3 & $180 \times 6$ & 9 & 9.8$\pm$3.4 & $-$35.9$\pm$4.4 \\
% & & & & & (7.7$\pm$2.8) & \\
%
%
% ID8---H5
& 5 & 1.1 & $180 \times 8$  & 46 & 5.7$\pm$0.7 & $-$13.7$\pm$0.9 \\
%&  & & & & (5.4$\pm$0.6) & \\
\hline
\end{tabular}
\tablecomments{$^{\rm a}$The pixel-to-pixel S/N of the continuum level.} 
\end{center}
%{Notes...\\
%De03: DeWarf+2003 }
%
\end{table*}
\begin{figure}[!h]
\begin{center}
% /Users/chikako/WORK/HImetal/QUARTET/fitsfile/ukidssQUARTET_Ks_CF_HAeBe0409.eps
\includegraphics[scale=0.55]{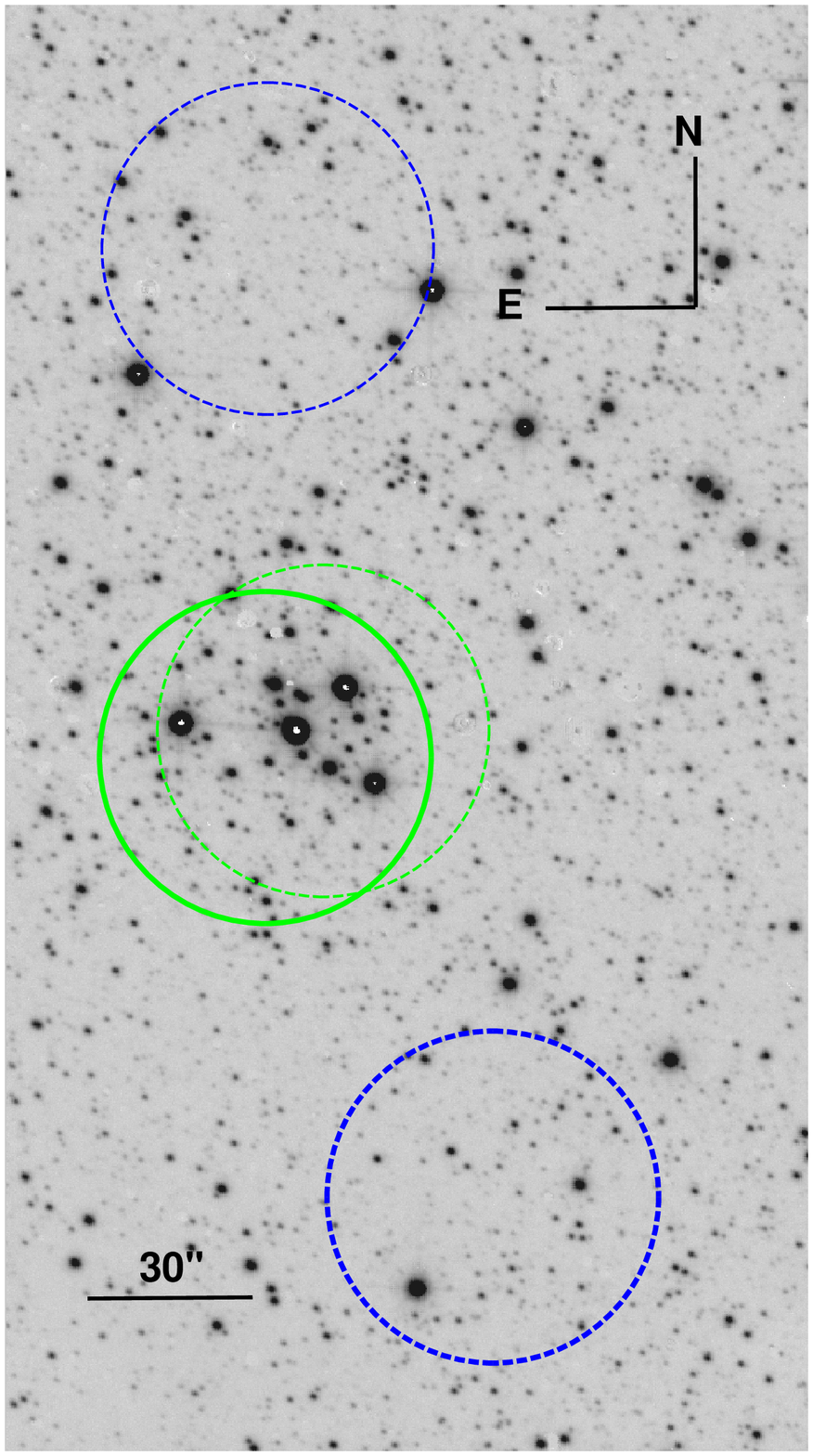} 
%
%
% x 3500--10000 logscale 
%
% scale: asinh,  limits: 99% 
% xpaset -p ds9 regions delete all 
% cat /Users/chikako/WORK/HImetal/QUARTET/fitsfile/QUA_HAeBe_2012Apr_RADec_point.reg | xpaset ds9 regions -format ds9 -system wcs
% cat /Users/chikako/WORK/IGaccretion/fitsfile/Qua_ID6_f12_RADec.reg | xpaset ds9 regions -format ds9 -system wcs
% # text(18:36:18.682,-7:04:45.54) textrotate=0 color=red font="helvetica 30 bold roman" text={H2}   
% /Users/chikako/WORK/HImetal/QUARTET/fitsfile/QUA_HAeBe_2012Apr_RADec_point0409_BW.eps
\includegraphics[scale=0.5]{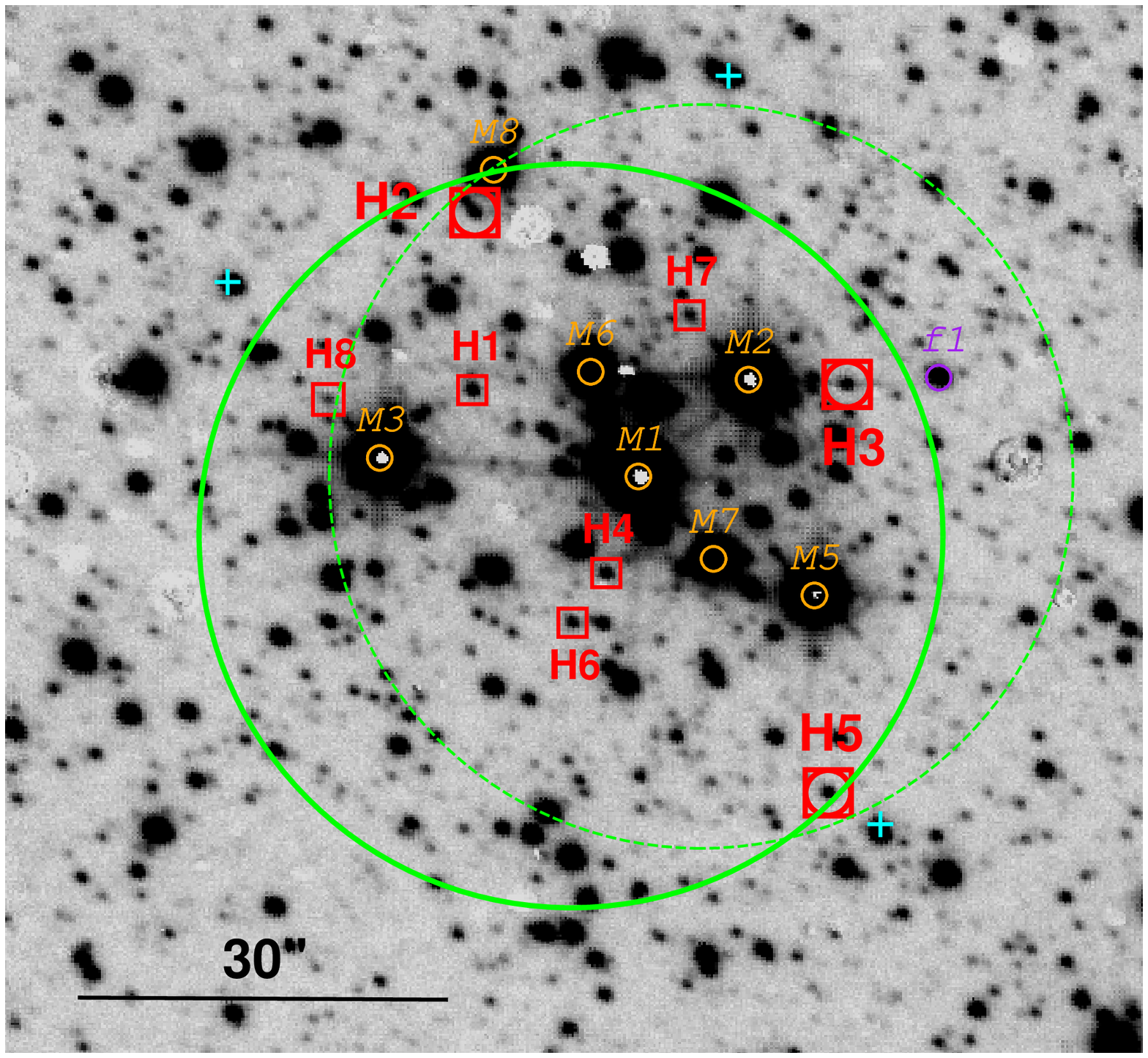}
\caption{UKIDSS Ks-band image of the Quartet cluster. 
Left: The green circle shows the cluster region, while blue dashed 
circles show the control field (the thick lower right circle for 
the main control field defined in Section~\ref{sec:results_img} 
and thin upper circle for the sub-control field defined in 
Section~\ref{sec:IMDF_Qua}).
Right: The close-up of the left figure around the cluster region. The
eight HAeBe candidates are shown with red boxes.
Number designations are from Tab.~\ref{tab:QUA_HAeBe}.
The three objects with spectroscopic observation are shown with
large red boxes. 
AO natural guide stars for IRCS spectroscopy are shown with aqua plus
 symbols.
The bright stars observed in \citet{Messineo2009} are shown with orange
circles with the numbers.
%
%Note that No. 4 and No. 9 stars are outside of the field (see the main
% text).
%
% % ID6---H3
In the IRCS spectroscopy of HAeBe 3, two stars, f1 (purple)
and the star No. 2 by \citealt{Messineo2009} ([MDI2009] Quartet 2) 
are also included in the same slit (see detail in
Appendix~\ref{sec:AppendixA}). 
An alternatively defined cluster region (Section~\ref{sec:IMDF_Qua}),
whose center is the centroid of very bright cluster members (No. 1, 2,
5, 6, and 7 in \citealt{Messineo2009}) and whose radius is the same as
the original cluster region, is shown with the green dashed circle.}
\label{fig:3col_img}
\end{center}
\end{figure}
%%%% xpaset -p ds9 file ukidssQUARTET_Ks.fits
% cat QUA_CF_2011Feb.reg | xpaset ds9 regions -format ds9 -system wcs
% cat /Users/chikako/WORK/HImetal/QUARTET/fitsfile/QUA_HAeBe_2012Apr_RADec.reg | xpaset ds9 regions -format ds9 -system wcs
%

%%%%% fig2 %%%%%
% cd /Users/chikako/WORK/HImetal/QUARTET/plfile
% Pro_QUA_phot2011Feb.pl
\begin{figure}[!h]
\begin{center}
% /Users/chikako/WORK/HImetal/QUARTET/plfile/Pro_QUA_phot2011Feb_woE.eps
\includegraphics[scale=0.5]{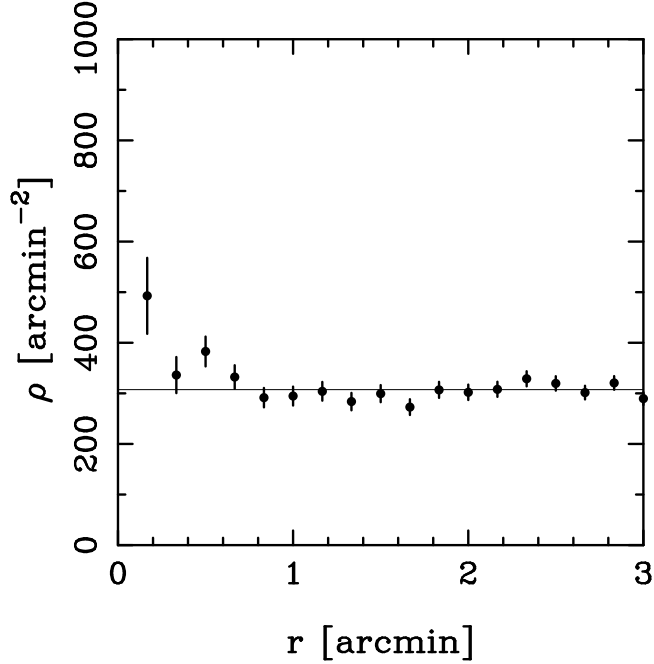}
%
% 1 sigma
\caption{Radial variation of the projected stellar density 
(filled circles) in the Quartet cluster with the center of 
$(\alpha_{\rm 2000}$, $\delta_{\rm 2000}) = (18^{\rm h} 36^{\rm m}
17.7^{\rm s}, -07^\circ 05' 12.1'')$.
The error bars represent Poisson errors.
The horizontal solid line indicates the density of background stars.}
\label{fig:Pro_Qua}
\end{center}
\end{figure}

%%%%% fig3-1 %%%%%
% JHJ_QUA_2011Feb.pl
\begin{figure}[!h]
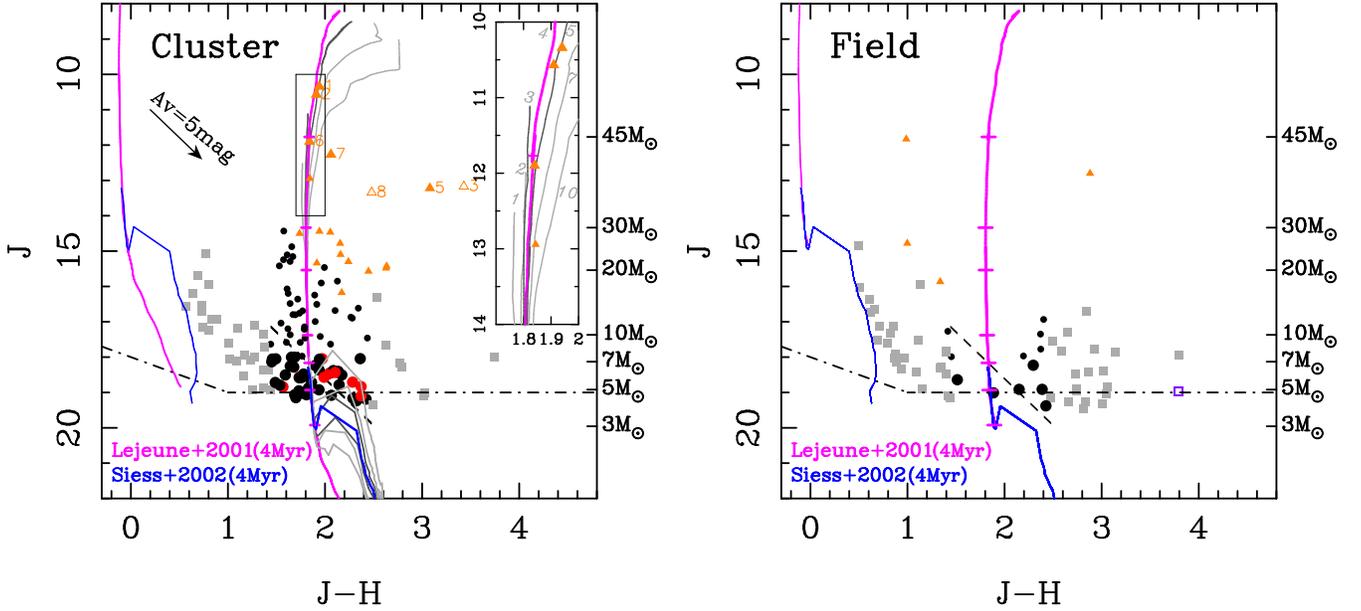

\begin{center}
% \includegraphics[scale=0.48]{/Users/chikako/WORK/HImetal/QUARTET/plfile/2012Apr/JHJ_QUA_2011Feb_Av1424_Ages2_cl_it.eps}
% /Users/chikako/WORK/HImetal/QUARTET/plfile/2012Apr/JHJ_QUA_2011Feb_Av1424_0421.eps
 \includegraphics[scale=0.48]{JHJ_QUA_2011Feb_Av1424_Ages2_cl_it.eps}
\hspace{0.5em}
\includegraphics[scale=0.48]{JHJ_QUA_2011Feb_Av1424_0813_con.eps}
\caption{(J-H) vs. J color-magnitude diagram of the Quartet cluster
region (left) and the control field (right).
Only stars detected with more than 10$\sigma$ in both J and H bands are
plotted. 
%
%The magenta lines show the dwarf tracks (Bessell & Brett, 1988) 
The magenta lines show the isochrone tracks by \citet{Lejeune2001}
($\ge$3\,$M_\odot$) for the age of 4\,Myr at the distance of 6\,kpc,
while the blue lines show the isochrone models by \citet{Siess2000}
($0.1 \le M/M_\odot \le 7$) for the same age at the same distance.
For both isochrone tracks with extinction of $A_V = 0$ and 18\,mag are
shown with thin and thick lines, respectively.
In the left figure, the isochrone tracks for the ages of 1, 2, 3, 5, 7,
and 10\,Myr are also shown with dark and light gray lines.
 The inset shows a blowup of the region enclosed with the black box. The
 gray italic numbers in the inset show the ages of isochrone models. 
The arrow shows the reddening vectors of $A_V = 5$\,mag.
The dot-dashed lines show the limiting magnitudes (10$\sigma$).
The short horizontal lines placed on the magenta lines show positions of
45, 30, 20, 10, 7, 5, and 3\,$M_\odot$, that are also shown on the right
y-axis.
Possible cluster members ($A_V = 14$--24\,mag) are shown with filled
circles, while the other stars are shown with gray squares.
Among the cluster members, IM stars ($\le$7\,$M_\odot$; $J \ge 18$\,mag)
are shown with large filled circles, red for stars with K-band excess
(HAeBe candidates) and black for stars without excess, while higher mass
stars are shown with small filled circles.
%
%Among them, stars with the K-band excess (HAeBe candidates) are shown
% with red, while those without the excess are shown with black (see the
% main text).
%
Three HAeBe candidates, which are targets for spectroscopic observation,
are shown with filled circles enclosed by open circles.
The brightest stars identified by \citet{Messineo2009} in the cluster
region are shown with large orange triangles, filled ones for possible
cluster members and open ones for field stars.
Bright stars in 2MASS Point Source Catalog in the cluster region and the
control field are shown with small orange triangles in left and right
figures, respectively. 
Additionally, the lines that pass through the isochrone tracks by
\citet{Lejeune2001} at the mass of 7\,$M_\odot$ and are parallel to the 
reddening vector are shown with thin dashed lines.
The f1 star (see Appendix~\ref{sec:AppendixA}) is shown with purple open
square in the right figure.} \label{fig:CM_Qua}
\end{center}
\end{figure}

%%%%% fig4 %%%%%
% .pl
\begin{figure}[!h]
\begin{center}
% /Users/chikako/WORK/HImetal/QUARTET/plfile/2012Apr/Av_dist_Qua_2011Feb_f2_4My.eps
% JHJ_QUA_2011Feb_Av1424_0813.pl
\includegraphics[scale=0.5]{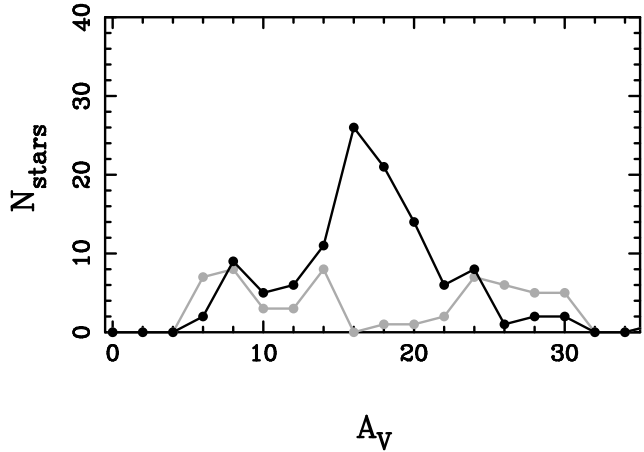}
\caption{$A_V$ distributions of the stars in the cluster region (black
line) and the control field (gray line).} 
\label{fig:Av_Qua}
\end{center}
\end{figure}

%%%%% fig5 %%%%%
% CC_QUA_2012Apr.pl
\begin{figure}[!h]
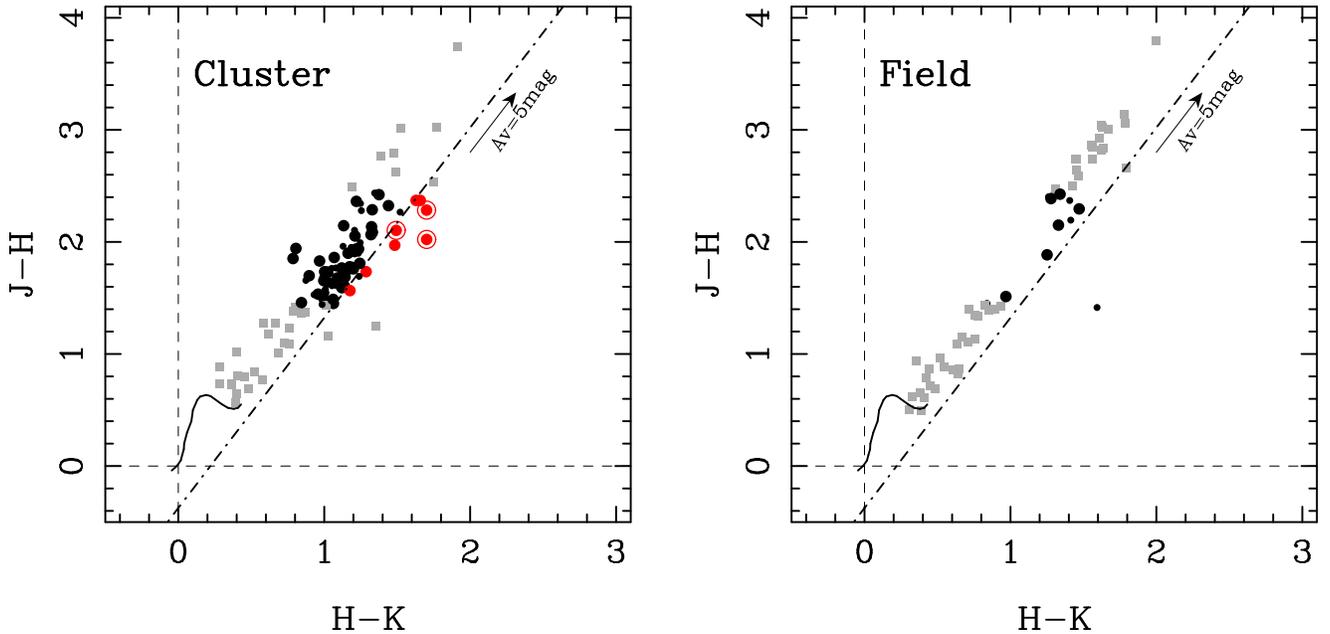

\begin{center}
% /Users/chikako/WORK/HImetal/QUARTET/plfile/2012Apr/CC_QUA_2012Apr_MKO_Av1424.eps
\includegraphics[scale=0.5]{CC_QUA_2012Apr_MKO_Av1424.eps}
\hspace{2em}
% /Users/chikako/WORK/HImetal/QUARTET/plfile/2012Apr/CC_QUA_2012Apr_MKO_Av1424_f2.eps
\includegraphics[scale=0.5]{CC_QUA_2012Apr_MKO_Av1424_f2.eps}

\caption{Color-color diagram of the stars in the Quartet cluster
region (left) and in the control field (right). 
Stars with extinction of $A_V = 14$--24\,mag are shown with filled
circles, while the other stars are shown with gray squares.
Among stars with extinction of $A_V = 14$--24\,mag, probable IM stars
($J \ge 18$\,mag) are shown with large filled circles, while higher mass
stars ($J<18$\,mag) are shown with small filled circles. 
Among them, stars with the K-band excess (HAeBe candidates) are shown 
with red, while those without the excess are shown with black (see
the main text).
Targets for spectroscopic observation are shown with filled
circles enclosed by open circles. 
Only stars detected with more than 10$\sigma$ in the J, H, and K bands
are plotted.
The estimated borderline in the MKO system, which is parallel to 
the reddening vector and distinguish HAeBe stars from other objects, 
is shown with dot-dashed lines.
The dwarf tracks in the MKO system \citep{Yasui2008} are shown with the
black lines.}
\label{fig:CC_Qua}
\end{center}
\end{figure}

%%% fig6 %%%
% ./G9_presen.pl
\begin{figure}[!h]
\begin{center}
%
% /Users/chikako/WORK/IGaccretion/plfile/Qua_H235_s2Fn_cont_wSTDfk.eps
\includegraphics[scale=0.43]{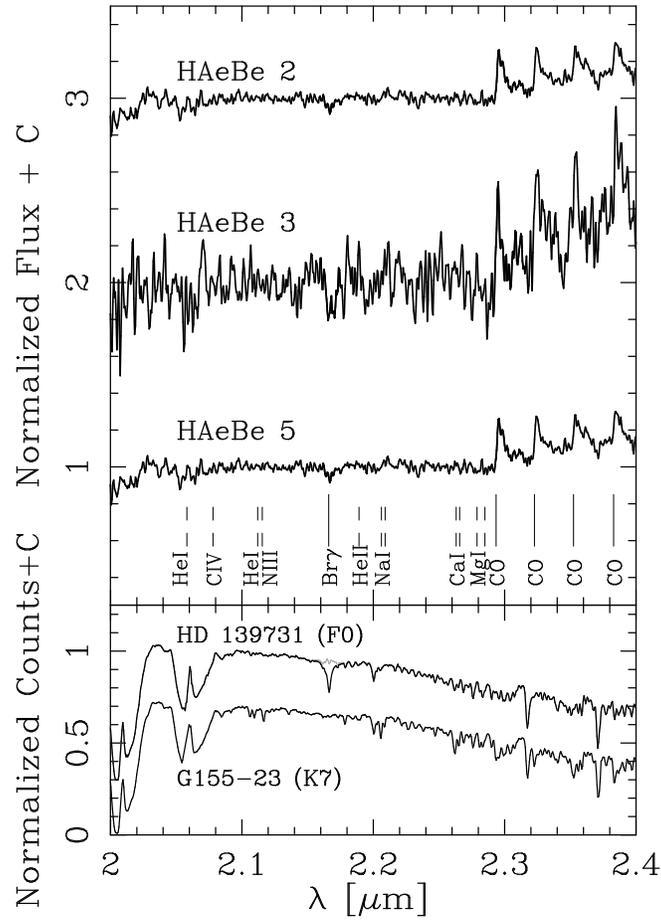}
\caption{IRCS K-band spectra of three HAeBe candidates in the Quartet
cluster (top) and those of two telluric standard stars, HD 139731 and
G155-23 (bottom).
In the top plot, the spectra have been telluric-corrected and normalized
to 1.0 after 2-pixel smoothing, and they are plotted with an offset for
clarity.
All objects show both Br$\gamma$ absorption and CO bandhead emissions. 
Detected and non-detected absorption/emission lines are shown with
vertical solid and dashed lines, respectively.
In the bottom plot, the spectra show the counts normalized to 1.0 at
$\lambda=2.1$\,$\mu$m, and they are plotted with an offset for clarity.
For HD 139731, the observed spectrum is shown with black, while that
after eliminating the Br$\gamma$ absorption line is shown with gray (see
detail in the main text).}
\label{fig:Spec_Qua}
\end{center}
\end{figure}

%%% fig7 %%%
% cd /Users/chikako/WORK/paper/HAeBe_TS/RefereeC/SurvivalAnalysis/STSDAS_asrv/newAGE
% SA_logIMDF_2014Jul_newAGE.pl

% cd /Users/chikako/WORK/HImetal/HAeBe_reference/plfile/FIT
% fitDF_agesum_2014Jan.pl

% /Users/chikako/WORK/HImetal/HAeBe_reference/plfile/FIT
% ./logIMDFsp_2014Jan3.pl 110
\begin{figure}[!h]
\begin{center}
% /Users/chikako/WORK/HImetal/HAeBe_reference/plfile/FIT/logIMDFsp_2014Jan3_QUA0814.eps
%\includegraphics[scale=0.6]{logIMDFsp_2014Jan3_QUA0814.eps}
% 
%\includegraphics[scale=0.6]{/Users/chikako/WORK/HImetal/HAeBe_reference/plfile/FIT/logIMDFsp_2015_Qua.eps}
% \includegraphics[scale=0.6]{/Users/chikako/WORK/HImetal/HAeBe_reference/plfile/FIT/HAeBeDF_2014Jan3_Qua.eps}
\includegraphics[scale=0.6]{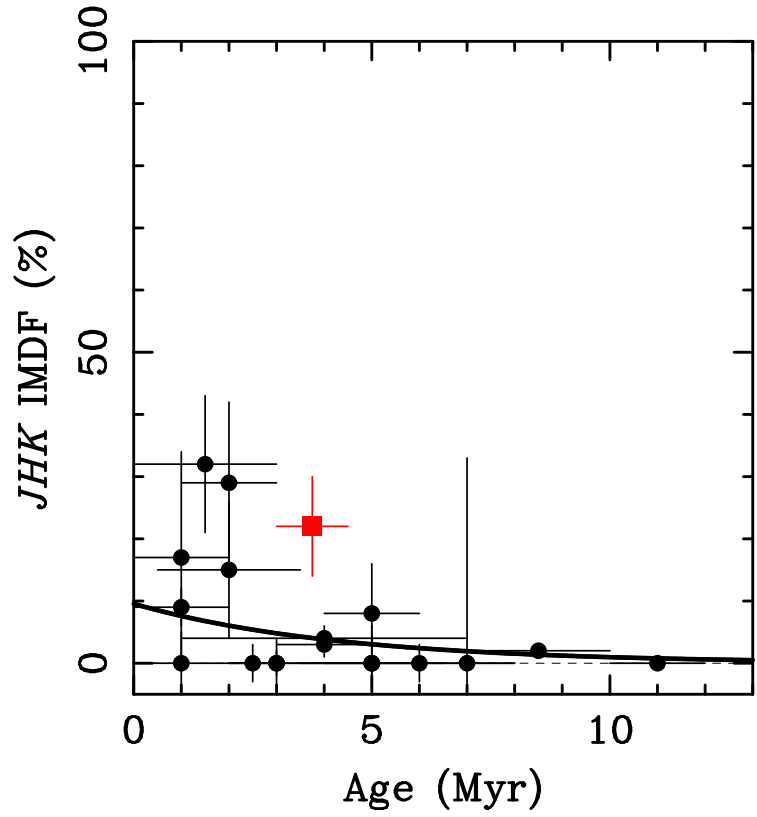}
\caption{Comparison of JHK intermediate-mass disk fraction (IMDF)
between the Quartet (red filled square) and young clusters in the solar 
neighborhood (black filled circles that are from the red points in
Fig. 5 (left) of \citealt{Yasui2014}).
The JHK IMDF vs. age relation is shown with the black line, which is
from red line in Fig.~5 (left) of \citet{Yasui2014}.
 It should be noted that the stellar mass range for the estimation of
 IMDF for the Quartet is on the higher-mass side ($\sim$5--7\,$M_\odot$)
 than those in the solar neighborhood ($\sim$1.5--7\,$M_\odot$) (see
 Section~\ref{sec:IMDF_Qua}).} \label{fig:IMDF_Qua}
\end{center}
\end{figure}

\clearpage

\appendix
\section{Additional data of IRCS spectroscopy}
\label{sec:AppendixA}

\begin{figure}[h]
\begin{center}
%
% /Users/chikako/WORK/IGaccretion/plfile/Qua_ID6f12_s2Fn_cont.eps
\includegraphics[scale=0.5]{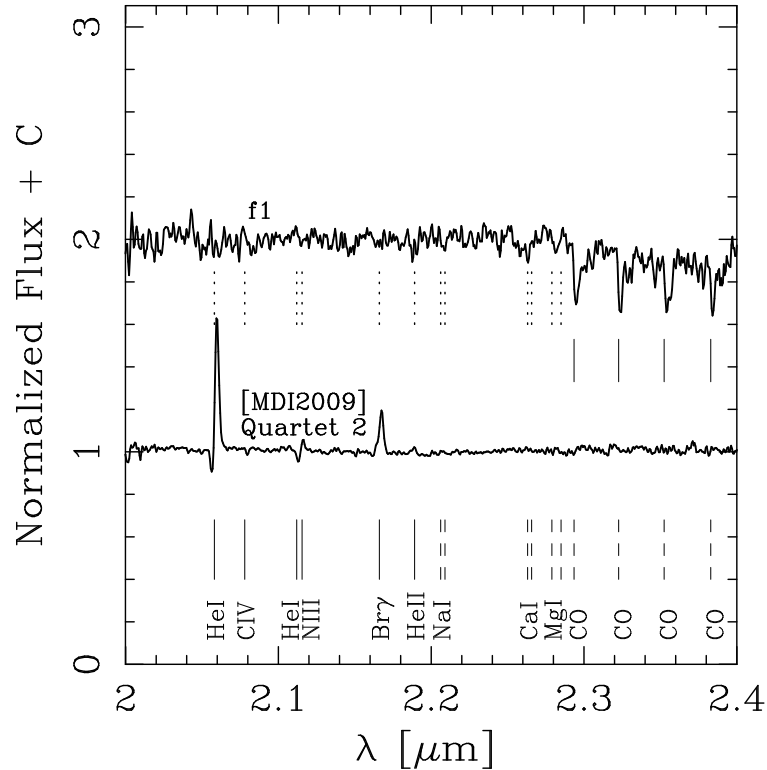}
\caption{IRCS K-band spectra of the f1 star and [MDI2009] Quartet 2, 
which are stars close to HAeBe 3.
Detected and non-detected lines are shown with vertical solid and dashed
lines, respectively.}
\label{fig:Spec_Quaf12}
\end{center}
\end{figure}

In the IRCS spectroscopy of HAeBe 3, two stars (the f1 star and
[MDI2009] Quartet 2) are also included in the same slit 
(see Fig.~\ref{fig:3col_img}). 
[MDI2009] Quartet 2 is the same star as the star No.2 in
\citet{Messineo2009}. 
The f1 star is located to the $\sim$7$''$ west of HAeBe 3, while 
[MDI2009] Quartet 2 is located to the $\sim$8$''$ east of HAeBe 3.
The K band magnitude of the f1 star is estimated to be 13.12\,mag 
in Section~\ref{sec:imging}, while that of the [MDI2009] Quartet 2 is 
7.58\,mag from 2MASS point source catalogue \citep{Messineo2009}.
% QUA_JHK_2011Feb5_1234qr.dat
% 4543    1776.85  3732.59 -7.463 0.066  17833   1777.97 3731.179 -11.607 0.043  0  6683    1780.13  3730.01 -12.184 0.060  0
% delK = 25.301 ---> -12.184+25.301 = 13.117
%
Because the f1 star has a large extinction ($A_V \sim 36$\,mag; 
%$J=18.977$, $H=15.19$; see Fig~\ref{fig:CM_Qua}, purple open 
square), it should not be a member but a background star.  
Because each star is included in the slit in only one position out 
of two by nodding (see Section~\ref{sec:spec}), integration time is 
9\,min ($180\,{\rm sec} \times 3$) for both stars.
Although the f1 star was almost centered on the slit, [MDI2009]
Quartet 2 was a little off center.
Spectra of the two stars are shown in Fig.~\ref{fig:Spec_Quaf12}.  The
S/N of the spectra of f1 and [MDI2009] Quartet 2 are 23 and 115,
respectively.
Considering the S/Ns, lines with equivalent width of $\ge$1.4\,\AA\ 
is detected with 3$\sigma$ for the f1 star, while that of 
$\ge$0.3\,\AA\ is detected for [MDI2009] Quartet 2.

The spectrum of the f1 star shows CO bandhead absorption, although
other lines are weak or not detected.
For luminosity class V stars, significant CO bandhead absorption lines
are detected in stars later than F-type although the Na absorption line
at 2.20\,$\mu$m is also detected comparably in those stars
\citep{Rayner2009}.
For luminosity class III stars (or earlier), the CO absorption lines are
detected in stars later than G-type and Na absorption is very small (EW
of $<$1.4\,\AA) in stars earlier than mid K-type \citep{Rayner2009}. 
Therefore, f1 is suggested to be an early K III star. 
The spectrum of [MDI2009] Quartet 2 shows \ion{He}{1} 2.1126\,$\mu$m
absorption, \ion{N}{3} 2.116\,$\mu$m emission, and Br$\gamma$ emission
lines, which were all detected by \citet{Messineo2009}. 
In addition, the \ion{He}{2} 2.188\,$\mu$m emission line is also
detected.

\end{document}